# Towards Effective Human-AI Collaboration in GUI-Based Interactive Task Learning Agents

**Toby Jia-Jun Li**
Carnegie Mellon University
tobyli@cs.cmu.edu

**Jingya Chen**
Carnegie Mellon University
jingyach@andrew.cmu.edu

**Tom M. Mitchell**
Carnegie Mellon University
tom.mitchell@cs.cmu.edu

**Brad A. Myers**
Carnegie Mellon University
bam@cs.cmu.edu

## Abstract
We argue that a key challenge in enabling usable and useful interactive task learning for intelligent agents is to facilitate effective Human-AI collaboration. We reflect on our past 5 years of efforts on designing, developing and studying the SUGILITE system, discuss the issues on incorporating recent advances in AI with HCI principles in mixed-initiative interactions and multimodal interactions, and summarize the lessons we learned. Lastly, we identify several challenges and opportunities, and describe our ongoing work.

## Introduction
Enabling end-users to automate their tasks using intelligent agents has been a long-standing objective in both the HCI and AI communities. A key research problem is enabling users to teach the agents new tasks. Despite the wide-adoption of existing agents like Siri, Google Assistant and Alexa, their capabilities are limited to domains that are either built-in or programmed by third-party expert developers. Prior studies have shown that users' automation needs are highly diverse and personalized, with a "long-tail" that is not currently supported by prevailing agents [7]. Therefore, supporting end user development (EUD) for task automation agents is particularly useful.

A promising approach towards this direction is to leverage the resource of existing graphical user interfaces (GUIs) of third-party apps. The GUIs encapsulate rich knowledge about the flows of the underlying tasks and the properties and relations of relevant entities, so they can be used to bootstrap the domain-specific knowledge needed by the agents without requiring pre-programmed prior knowledge in specific task domains [11]. Users are also familiar with GUIs, which makes GUIs the ideal medium to which users can refer during task instructions [8,10].

Significant progress has been made on this topic in recent years in both AI and HCI. Specifically on the AI side, advances in natural language processing (NLP) enable the agents to process users' instructions of task procedures, conditionals, concepts definitions, and classifiers in natural language [2,6,10], to ground the instructions (e.g., [12]), and to have dialog with users based on GUI-extracted task models (e.g., [11]). Reinforcement learning techniques allow the agent to more

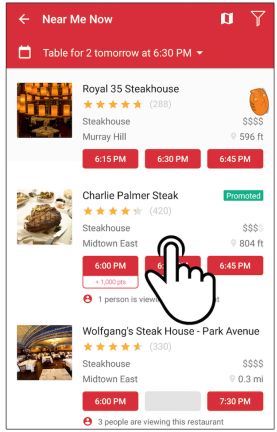

a. User demonstrates the action directly on unmodified GUIs of third party apps

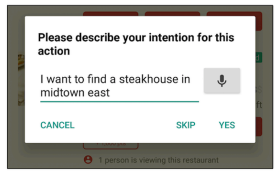

b. SUGILITE asks the user to describe intentions for actions

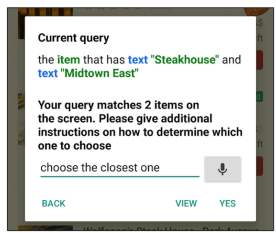

c. Multi-turn conversations help users refine ambiguous descriptions

effectively explore action sequences on GUIs to complete tasks [13]. Large GUI datasets such as RICO [4] allow the analysis of GUI patterns at scale, and the construction of generalized models for extracting semantic information from GUIs.

The HCI community also has presented new study findings, design implications, and interaction designs in this domain. A key direction has been the design of multimodal interfaces that leverage both natural language instructions and GUI demonstrations [1,7]. Prior work also explored how users naturally express their task intents [10,15,17] and designed new interfaces to guide the users to provide more effective inputs (e.g., [8]).

We argue that a key problem in this domain is to facilitate effective Human-AI collaboration in the interactive task learning (ITL) process. On one hand, AI-centric task flow exploration and program synthesis techniques often lack transparency for users to understand the internal process, and they provide the users with little control over the task fulfillment process to reflect their personal preferences. On the other hand, machine intelligence is desired because the users' instructions are often incomplete, vague, ambiguous, or even incorrect. Therefore, the system needs to provide adequate assistance to guide the users to provide effective inputs to express their intents, while retaining the users' agency, trust, and control of the process. While relevant design principles have been discussed in early foundational works in mixed-initiative interaction [5] and demonstrational interfaces [16], incorporating these ideas into the design and implementation of actual systems remains an interesting challenge.

In this position paper, we first summarize the lessons we learned from designing, implementing, and studying the SUGILITE agent in the past five years. We then identify several challenges and opportunities in this field, and describe our ongoing work in these areas.

## SUGILITE Overview

SUGILITE is a smartphone-based interactive task learning agent that enables the users to teach new tasks and relevant concepts using a combination of natural language instructions and app GUI demonstrations. It presents several interesting features such as the use of GUIs to ground and to parameterize language instructions [7,10], the use of interactive *mutual disambiguation* to clarify demonstrations and natural language instructions [8], the use of app GUIs as the medium to invoke and read data from IoT devices [9], and the generalization of learned concepts across different task domains [10]. See the individual papers for the detailed descriptions of these features.

## Lessons Learned

*Studying the User's <u>Natural</u> Programming Style*
We found that a crucial step in the design process is to understand how users *naturally* instruct the tasks, explain the relevant concepts, and express their intents. When users interact with an agent, "code-switch" often occurs, where the users adjust the styles and contents of their expressions to adapt to their expectations of the system's capability [3]. This phenomenon is not helpful in our design process, because the users' expectations are based on their prior experience interacting with the prevailing agents, while we are trying to develop a new system with new capabilities beyond the prevailing ones.

For example, during the development of SUGILITE's concept instruction framework, we conducted a formative study (details in [10,18]) on how users naturally instruct task conditionals, and how mobile app contexts influence their instructions. We specifically asked them to not consider the technical limitations of the system, and used the Natural Programming Elicitation method [17], where we showed graphical representations of the tasks with limited text in the prompts to reduce biases in user responses. The results helped us understand that (1) users frequently used ambiguous, unclear or vague concepts, in the instructions; (2) they also often expected the system to have the capability of commonsense reasoning with world knowledge; and (3) simply *seeing* the GUI context of the underlying apps

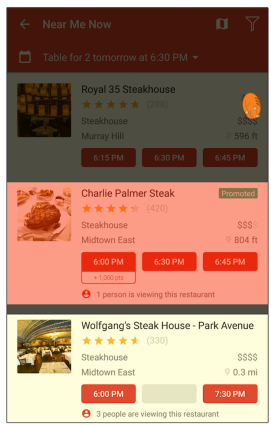

d. User can view the result for the current query and the originally clicked UI object

Figure 1: The screenshots of SUGILITE's demonstration mechanism and its multi-modal mixed-initiative intent classification process for the demonstrated actions.

reduced their usage of ambiguous, unclear or vague concepts, and made them refer to content on the GUI more often.

*Promoting System Initiatives to Guide User Inputs*

We found that a key challenge for the users of an EUD agent is to understand: (1) what can be done; (2) what "building blocks" are available; and (3) what strategies can be used to express their intents with the available building blocks. The answers to these questions are especially non-obvious in natural language agents, which leads to frequent breakdowns in conversations [3]. The users' initial task intents are also often uncertain and vague in nature, and need the agent's help to refine and clarify them.

Referring the users to concrete examples that they are familiar with based on the agent's *guess* of the user's intent can be helpful [8]. For example, as shown in Figure 1, when the user demonstrates an action of selecting an item, the agent needs to understand why the user selected this item so that it can generalize the learned procedure to perform the task in different task scenarios. SUGILITE'S approach is to ask the user to *verbally* explain why they selected this item, and visualize the query translated from the user's explanation on the GUI through an interactive overlay. If the query does not match the demonstrated action, the user can refine the instruction with the help of the visualization. If the query is ambiguous (i.e., matches the demonstrated item in addition to some false positives), the overlay highlights the correct match and the false positives in different colors, and asks the user to focus on explaining the key differences between them. Our study found this mechanism to be effective in helping users refine their data description instructions to accurately reflect their intents [8].

We used a similar strategy in designing SUGILITE's concept instruction framework, where the agent allows the use of ambiguous, vague, or unknown concepts in verbal explanations, and then recursively resolves them with the user, and proactively prompts the user to refer to app GUIs during the concept resolution when opportunities arise (see [10] for details).

## Challenges and Opportunities

*Extracting Task Semantic from GUIs*

SUGILITE illustrates the promise of using GUIs as a resource for grounding natural language instructions. A major challenge in natural language instruction is that the users do not know what concepts the agent already knows that they can use in their instructions [10]. Therefore, they often introduce additional unknown concepts that are either unnecessary or beyond the capability of the agent (e.g., explaining "hot" as "when I'm sweating" in "open the window when it is hot"). By using the app GUIs as the medium, the system can effectively constrain the users to refer to things that can be found out from some app's GUI (e.g., "hot" means "the temperature is high"), which mostly overlaps with the "capability boundary" of smartphone-based agents, and allows the users to define new concepts for the agent by referring to app GUIs [7,10].

An interesting future direction is to better extract *semantics* from app GUIs so that the user can focus on high-level task specifications and personal preferences without dealing with low-level mundane details (e.g., "buy 2 burgers" means setting the value of the textbox below the text "quantity" and next to the text "Burger" to be "2"). Some research has made progress in this domain [14] thanks to the availability of large datasets of GUIs [4]. Recent reinforcement learning-based approaches and semantic parsing techniques have also shown promising results in learning models for navigating through GUIs for user-specified task objectives [13]. For task learning, an interesting challenge is to combine these user-independent domain-agnostic machine-learned models with the user's personalized instructions for specific tasks. This will likely require a new kind of mixed-initiative instruction [5] where the

| | |
|---|---|
| **Wrong dialog frame** | Responding "what kind of cuisine would you like" for the command "find me a place to sleep in Chicago tonight." |
| **Extract the wrong parts in the input as entities** | Extracting "Singapore" as the departure city in "Show me Singapore Airlines flights to London" |
| **Link the extracted phrase to wrong entities in its knowledge base** | Resolving "apple" in "What's the price of an apple" as the entity "Apple Inc." and therefore invoking the stock price lookup frame instead of a grocery frame |

Table 1: Examples of intent classification and entity recognition errors that we hope to be able to handle in our future work.

## Acknowledgement

This research was supported in part by Oath and Verizon through the InMind project, a JP Morgan Faculty Research Award, NSF grant IIS-1814472, and AFOSR grant FA95501710218. Any opinions, findings or recommendations expressed here are those of the authors and do not necessarily reflect views of the sponsors.

agent is more proactive in guiding the user and takes more initiative in the dialog. This could be supported by improved knowledge and task models, and more flexible dialog frameworks that can handle the continuous refinement and uncertainty inherent in natural language interaction, and the variations in user goals.

*Interfaces for Conversational Breakdown Repairs*
Another opportunity for applying HCI techniques in this domain is to help end users identify, handle, and recover from conversational breakdowns in their interactions with the agent. Specifically, our ongoing work focuses on errors from two key components in the agent's natural language understanding pipeline: *intent classification* and *entity recognition.* Intent classification errors are those where the system misrecognizes the intent in the user's utterance, and subsequently invokes the wrong dialog frame (examples shown in Table 1). Similarly, in entity extraction errors, the system either extracts the wrong parts of the input as entities or links the extracted phrase to the wrong entities in its knowledge base.

We are particularly interested in exploring the use of multi-modal interfaces to address these problems. We are current designing new interfaces where the user can refer to relevant apps, screens within apps, and GUI elements on screens when trying to explain their observed errors and fix the issues in natural language. We envision that this technique can enable users to provide concrete relevant (both positive and negative) examples, and prompt them to explain *how* each example relates to their underlying task intent and the errors that the agent made in the conversation.